\documentclass[epj,nopacs]{svjour}
\usepackage{graphicx}

\begin{document}

\title{Identification and Rejection of Fake Reconstructed Jets From a Fluctuating Heavy Ion Background in ATLAS}

\author{N.~Grau\inst{3} \and B.~A.~Cole\inst{3} \and W.~G.~Holzmann\inst{3} 
\and M.~Spousta\inst{2} \and P.~Steinberg\inst{1} (for the ATLAS 
Collaboration)}

\institute{
Brookhaven National Laboratory, Upton, New York, USA
\and Charles University, Faculty of Mathematics and Physics, Prague, Czech Republic
\and Columbia University, New York, New York, USA
}

\authorrunning{N.~Grau {\it et al.}}
\titlerunning{Identification and Rejection of Fake Reconstructed Cone Jets}

\date{Received: date / Revised version: date}

\abstract{
Full jet reconstruction in relativistic heavy ion collisions provides new and 
unique insights to the physics of parton energy loss. Because of the large 
underlying event multiplicity in $A+A$ collisions, random and correlated 
fluctuations in the background can result in the reconstruction of fake jets. 
These fake jets must be identified and rejected to obtain the purest jet 
sample possible. A large but reducible fake rate of jets reconstructed using 
an iterative cone algorithm on HIJING events is observed. The absolute rate of 
fake jets exceeds the binary-scaled $p+p$ jet rate below 50 GeV and is not 
negligible until 100 GeV. The variable $\Sigma j_{T}$, the sum of the jet 
constituent's $E_{T}$ perpendicular to the jet axis, is introduced to identify 
and reject fake jets at by a factor of 100 making it negligible. This variable 
is shown to not strongly depend on jet energy profiles modified by energy 
loss. By studying azimuthal correlations of reconstructed di-jets, the fake 
jet rate can be evaluated in data.
}

\maketitle

\section{Introduction}\label{sec:intro}

Complete jet reconstruction in relativistic heavy ion collisions is essential 
for a direct measurement of the energy lost by a high-$p_T$ colored parton 
traversing the colored medium produced in these collisions. Measurements of 
such energy loss have been made from experiments at Brookhaven National 
Laboratory's (BNL) Relativistic Heavy Ion Collider (RHIC) where single 
particle production rates in $\sqrt{s_{_{NN}}}$ = 200 GeV $Au+Au$ collisions 
are suppressed relative to binary-collision scaled $p+p$ rates at the same 
$\sqrt{s}$ \cite{PHENIXpi0s}. These measurements alone, however, are 
insufficient to constrain the details of the energy loss mechanism. For 
example, the current data cannot yet be used to determine the relative 
contributions of radiative and collisional enery loss, or even determine 
whether jet quenching is a perturbative or non-perturbative process. Because 
of the large underlying event multiplicity, dN/d$\eta\sim$ 600 for the top 
10\% $Au+Au$ collision cross section \cite{PHOBOSdNdeta}, direct jet 
reconstruction at RHIC is difficult, though attempts are underway 
\cite{Putschke:2008wn}\cite{Lai:2008zp}. As a proxy for (di-)jet 
reconstruction, 
two, high-$p_T$ particle azimuthal correlations have been measured. Results 
from two-particle correlations reveal that the trigger jet is essentially 
unmodified while the recoil jet is strongly suppressed \cite{Adler:2002tq}, 
further lending support to the idea that parton energy loss is the mechanism 
of the suppression. However, two-particle correlations have not revealed 
expected modifications to jet properties \cite{STARjets} such as softening of 
the fragmentation function \cite{SoftFF}, broadening of the di-jet 
acoplanarity \cite{DijetBroadening}, etc.~expected from energy loss models 
that reproduce the single particle suppression. One possible reconciliation 
between the single particle and two-particle results is that the two-particle 
results are ``energy loss biased''. That is, by requiring that two high-$p_T$ 
particles be present, events are chosen where the jet loses little, if any, 
energy. This can occur when the jet traverses a short path 
length \cite{PQM} or emits no gluons due to fluctuations in radiation 
(punchthrough).

Direct measurements of jets reconstructed via standard algorithms employed in 
$p+\bar{p}(p)$ or $e^{+}e^{-}$ collisions modified to handle the 
underlying heavy ion event, should reduce such energy loss biases. If the 
energy loss is dominated by gluon bremsstrahlung  radiated predominantly 
inside the jet cone \cite{Salgado:2003rv}, the lost energy will be recovered 
by reconstructing the full jet profile. If, however, energy loss is 
non-perturbative \cite{Kharzeev:2008qr} or radiates energy outside of the cone 
\cite{Lokhtin:2007ga}, a suppressed jet rate compared to binary-scaled $p+p$ 
will be observed. Whatever the scenario, direct measurements of jets will make 
a significant impact in understanding parton energy loss.

In the very near future the Large Hadron Collider (LHC) at CERN will deliver 
collisions of $Pb+Pb$ at $\sqrt{s_{_{NN}}}$ = 5.5 TeV. At this energy and 
nominal delivered luminosity, 20 million jets with $E_{T}>50$ GeV will be 
produced per month of $Pb+Pb$ running \cite{Accardi:2002vt}. These jets should 
be visible above the large underlying event multiplicity and can be 
reconstructed by jet reconstruction algorithms appropriately modified to 
handle the underlying event.

The ATLAS experiment will carry out extensive measurments of jet production in 
heavy ion collisions using its nearly hermetic electromagnetic and hadronic 
calorimeters and its large acceptance silicon tracking system. Techniques have 
been developed for iterative cone jet reconstruction in heavy ion collisions 
in ATLAS~\cite{Grau:2008ef}. These techniques require subtraction of the 
underlying event energy within the calorimeter prior to running the iterative 
cone jet algorithm (see Section~\ref{sec:jetrec}). Fluctuations in the 
background remain after subtraction and a stable cone, defined by the 
algorithm, can be obtained from random or correlated background. In 
studying any aspect of jets and their modification, such fake jets must be 
tagged and removed from the data sample. It is, therefore, necessary to study 
the characteristics of fake jets so as to extract a sample of jets with high 
purity.

This article describes the techniques that have been developed to identify and 
reject fake jets with the ATLAS detector arising from fluctuations in the 
underlying heavy ion event. The article is outlined as follows. In 
Section~\ref{sec:atlas} the relevant aspects of the ATLAS detector are 
defined. Section~\ref{sec:jetrec} presents the necessary details of jet 
reconstruction in heavy ion collisions in ATLAS and the identification of fake 
jets. This is followed by Section~\ref{sec:fakejets} where the results of fake 
jet detection are presented. A discussion of potential biases associated with 
the method of rejection and evaluation of rejection with data is given in 
Section~\ref{sec:results}.

\begin{figure}[t]
\includegraphics[width=\linewidth]{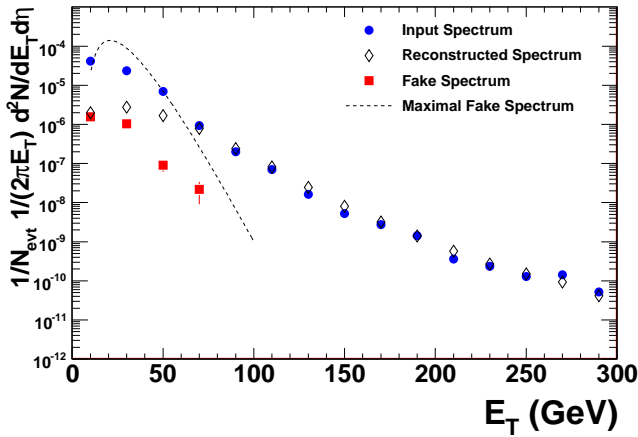}
\caption{Comparison of the spectrum of jets in dN/d$\eta$=2700, b=2 fm, 
HIJING events. The (blue) filled circles are the expected spectrum from 
binary-scaled $p+p$, the diamonds are the raw reconstructed 
spectrum uncorrected for jet reconstruction efficiency and energy 
resolution. The dashed line indicates the raw fake rate evaluated 
from hard-cut HIJING events. The (red) squares indicate the 
irreducible fake spectrum after rejection of the fake jets.}
\label{fig:spectrum}
\end{figure}

\section{The Atlas Detector}\label{sec:atlas}

The ATLAS detector \cite{ATLASTDR} is a large, multi-purpose, high energy 
physics experiment designed to detect rare processes in $p+p$ collisions at 
$\sqrt{s}$ = 14 TeV at the LHC. ATLAS has also been shown to be an 
excellent detector for measurements of both high $p_T$ and soft observables in 
heavy ion collisions \cite{ATLASHILoi}. 
The jet reconstruction described here is purely calorimetric. 
The ATLAS calorimeter is 
largely divided into a liquid argon (LAr) barrel electromagnetic calorimeter 
covering $|\eta|<$ 3.2, a LAr forward hadronic calorimeter covering $3.2<|\eta|<5$ 
and a barrel tile hadronic calorimeter covering $|\eta|<$ 1.5. Each 
calorimeter compartment is divided into longitudinal sections providing energy 
readout at different depths in the calorimeter. The barrel electromagnetic 
calorimeter is divided into 3 layers. The front ``strip'' layer, with 
approximately 6 $X_{0}$, has a typical cell (single readout) segmentation of 
$\Delta\eta\times\Delta\phi$ of 0.003$\times$0.1. The middle layer, with 24 
$X_{0}$, has a cell size of 0.025$\times$0.025 in 
$\Delta\eta\times\Delta\phi$. The barrel tile calorimeter has three 
layers with depth-dependent cell size; the first layer has 0.1$\times$ 0.1 segmentation. For reconstructing jets, towers in a 
fixed grid of 0.1$\times$0.1 in $\Delta\eta\times\Delta\phi$ are built from 
sums of cell energies in all layers.

\begin{figure*}
\begin{center}
\includegraphics[width=0.45\linewidth]{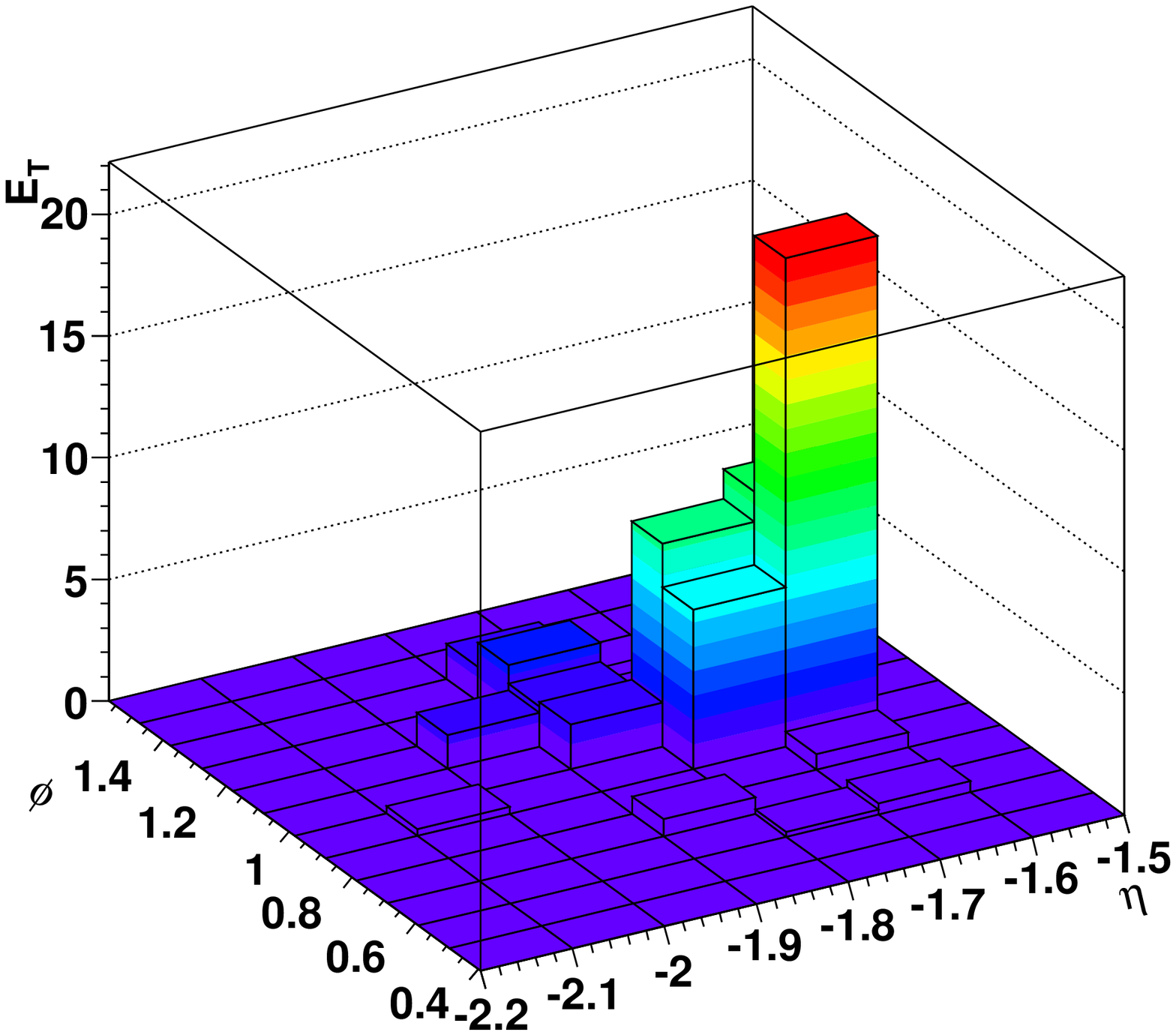}
\includegraphics[width=0.45\linewidth]{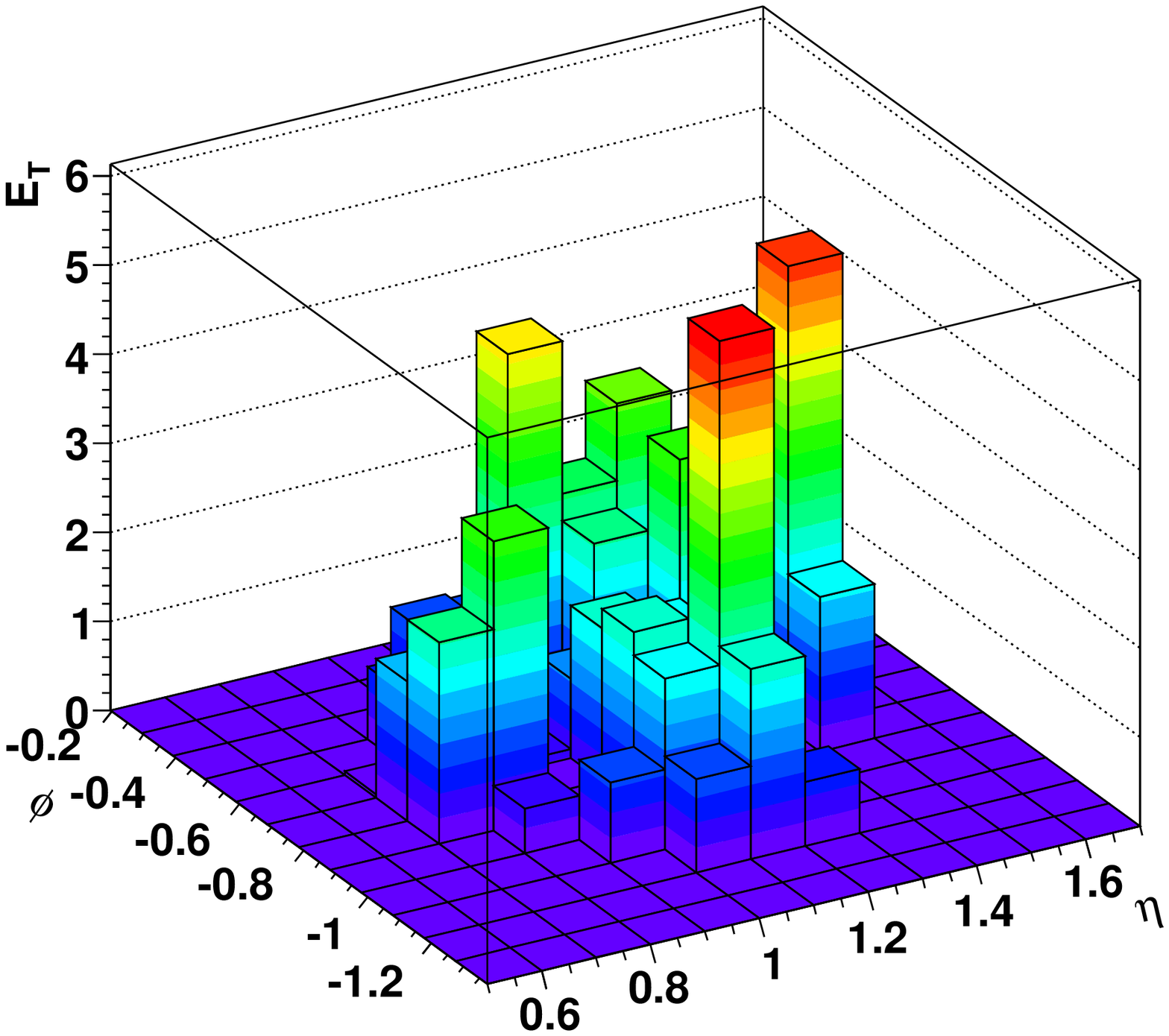}
\end{center}
\caption{Reconstructed jets with $E_T$=57 GeV from PYTHIA embedded into 
unmodified HIJING (left) and from hard-cut HIJING (right). The jet from the 
hard-cut HIJING is fake since no jets of this energy should exist in this 
sample (see text).}
\label{fig:jet}
\end{figure*}

\section{Jet reconstruction and fake jet identification}\label{sec:jetrec}

To evaluate the expected jet reconstruction performance in heavy ion collision 
with the ATLAS detector, di-jet PYTHIA \cite{Sjostrand:2006za} events were 
embedded in HIJING \cite{Gyulassy:1994ew} heavy ion events. The HIJING events 
(referred to as ``unmodified'') were generated without jet quenching and 
{\it without} a maximum $p_T$ cut on the hard scattering processes. The 
unmodified events result in correlated and fluctuating backgrounds from 
(mini-)jets, $c-\bar{c}$ and $b-\bar{b}$ production, and longitudinal string 
radiation all unsuppressed from the lack of quenching. The HIJING events 
containing embedded jets were propagated through the ATLAS detector using the 
GEANT4 software package, digitized to resemble raw data, and then processed by 
the standard ATLAS offline software chain.

Jet reconstruction was performed using a seeded iterative cone jet 
algorithm~\cite{Sterman:1977wj}\cite{OpalConeJets} with a seed tower energy of 
5 GeV and a cone radius $R = \sqrt{\Delta\phi^{2} + \Delta\eta^{2}} = 0.4$. On 
average, in unmodified HIJING events with dN/d$\eta$=2700, 150 GeV of 
background event $E_T$ is contained within the cone. The method used to 
subtract the background is to 
\begin{enumerate}
\item find regions of high energy in the calorimeter (potential jet regions)
\item calculate the average background $E_T$ in the calorimeter for each 
longitudinal layer $i$ and as a function of $\eta$, $E_{T}^{bkgr}\left(i,
\eta\right)$, excluding the regions found in step 1
\item subtract $E_{T}^{bkgr} \left(i,\eta\right)$ from all cells
\item build the 0.1$\times$0.1 grid of calorimeter towers from these modified 
cells. 
\end{enumerate}
The subtracted towers from this procedure are provided as input to the seeded 
cone jet algorithm.

Figure \ref{fig:spectrum} shows the invariant yields of jets in b = 2 fm, 
dN/d$\eta$ = 2700, HIJING events. The solid (blue) circles indicates the 
binary-scaled $p+p$ jet rate obtained from PYTHIA. The diamonds show the 
reconstructed spectrum, including embedded PYTHIA jets and fake jets which 
survive fake jet rejection (see Section~\ref{sec:fakejets}). The reconstructed 
spectrum is uncorrected for efficiency, energy scale, and energy resolution. 
The need for these corrections is clear below 70 GeV while the reconstructed 
and expected spectra agree well above 70 GeV. For these high-multiplicity 
HIJING events, 70 GeV jets have a reconstruction efficiency of 70\%, an energy 
scale offset of -2.5\%, and an energy resolution of 25\%. A complete 
evaluation of the centrality and $E_T$-dependence of the reconstruction is 
given in Ref.~\cite{Grau:2008ef}.

Fake jets were produced by generating, simulating in ATLAS, and performing jet 
reconstruction on a second set of HIJING events, labeled ``hard-cut''. These 
events differ from the unmodified HIJING events in that the parton spectrum at 
$Q^2 > 100$ GeV$^2$ is not sampled. Therefore, jets with $E_T > $ 10 GeV are 
purposefully suppressed with the intent being that any jet with $E_{T} \gg$ 10 
GeV is a fake jet. Even with a hard cut, jets above 10 GeV can occur from 
initial and/or final state radiation. What was also found was that 
longitudinal strings can radiate into midrapidity resulting in $10^{-2}$ jets 
with $E_{T}>$ 50 GeV per event. These were removed from the fake jet sample by 
excluding jets which match a HIJING parton above an appropriate threshold.

The resulting reconstructed jets from the hard-cut HIJING events then define 
the ``maximal'' fake rate plotted in Figure \ref{fig:spectrum}. The rate of 
fake jets exceeds that of the expected jet yield at low-$E_T$. The ratio of 
fake to reconstructed yields is 4.2 at 50 GeV, 0.33 at 70 GeV, and 0.03 at 90 
GeV. This shows that the contribution of fake jets decreases dramatically with 
reconstructed jet $E_T$. However, if it is desired to measure jets at moderate 
$E_T$, a rejection procedure is needed.

It is possible that the fake rate is overestimated. Part of the hard particle 
production in HIJING results from longitudinal strings, color fields between 
the nuclei, which, because of their large energy at the LHC, can fragment into 
a jet at midrapidity. Although an attempt was made to remove such jets from 
the hard-cut events, a sample of hard-cut HIJING events with the longitudinal 
string radiation suppressed was also studied. The maximal rate was reduced 
such that the ratio of the fake to reconstructed yield is 1.5 at 50 GeV, 0.11 
at 70 GeV, and 0.01 at 90 GeV. The longitudinal radiation, however, did not 
change the characteristics of the fake jets that were produced, nor their 
subsequent identification and rejection.

\begin{figure}[b]
\includegraphics[width=\linewidth]{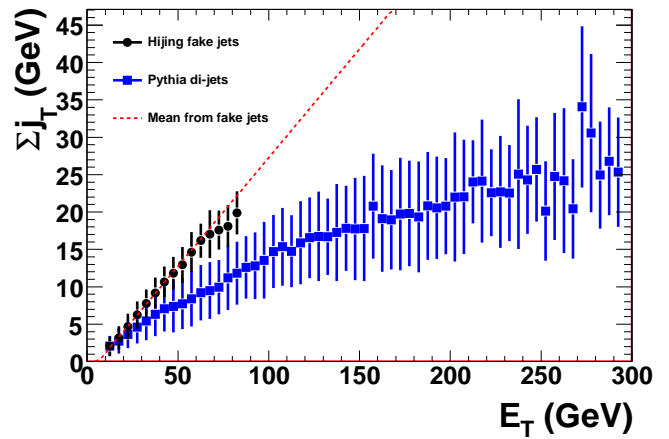}
\caption{The reconstructed jet $E_T$ dependence of $\Sigma j_T$ for fake jets 
from the unmodified HIJING sample (circles) and for reconstructed embedded 
jets in dN/d$\eta$=2700 (squares). The bars indicate the root-mean-square of 
the distribution in each $E_T$ bin. The dashed line is a fit to the HIJING 
fake jets. A clear separation between the fake jets and embedded jets is seen 
over all $E_T$. The larger $\Sigma j_T$ for the fake jets indicates that they 
have a broader energy distribution than the embedded jets.}
\label{fig:sumjtetdep}
\end{figure}

\begin{figure*}
\includegraphics[width=0.5\linewidth]{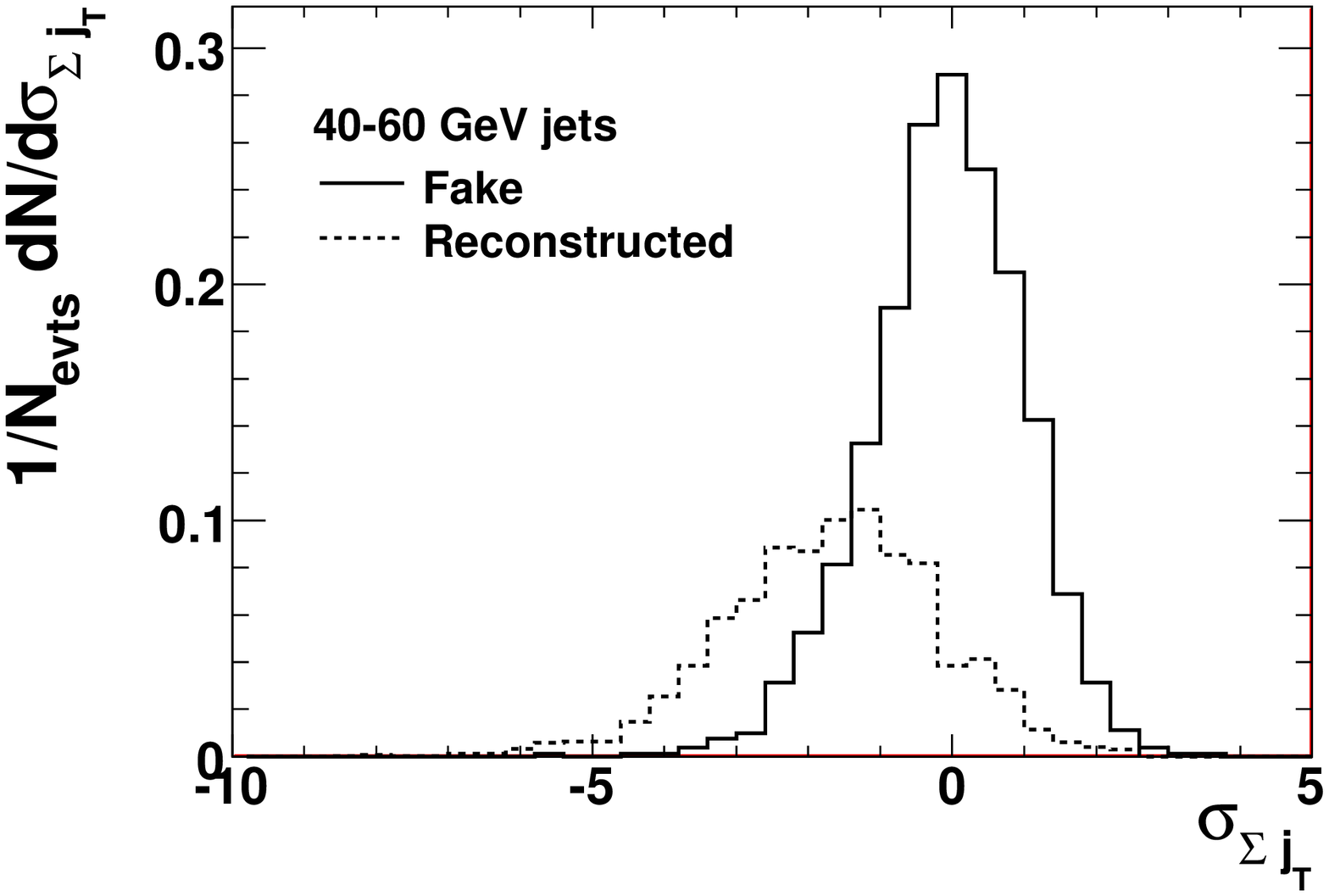}
\includegraphics[width=0.5\linewidth]{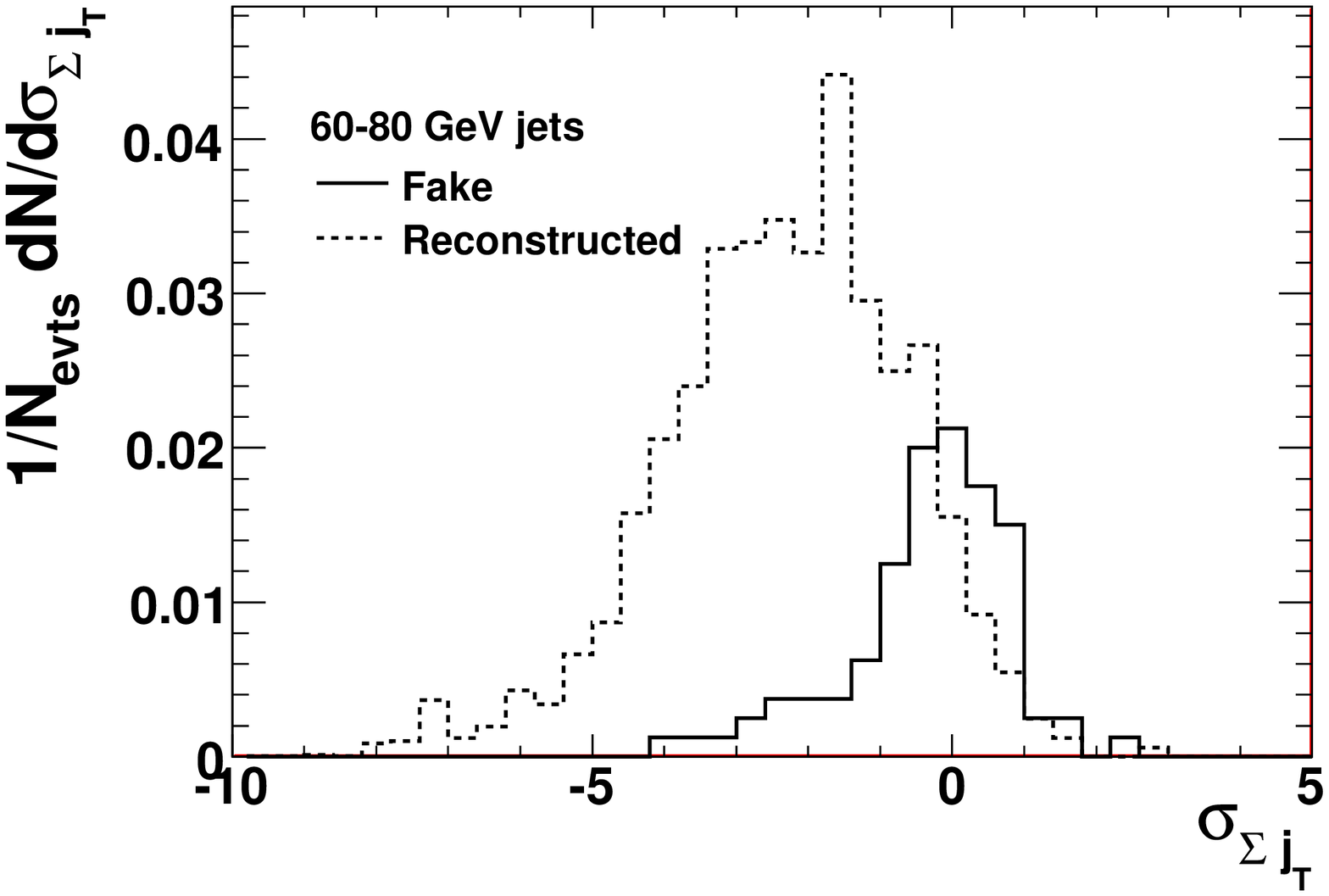}
\caption{The jet $E_T$-independent cut variable $\sigma_{\Sigma j_T}$, the 
number of $\sigma$ from the fake jet mean (see Eqn.~\ref{eq:sigmasumjt}) for 
two different jet $E_T$ bins 40-60 GeV (left) and 60-80 GeV (right). There is 
clear separation between the embedded jets (dashed line) compared to the fake 
jets (solid line).}
\label{fig:sigmasumjt}
\end{figure*}

\begin{figure}
\includegraphics[width=\linewidth]{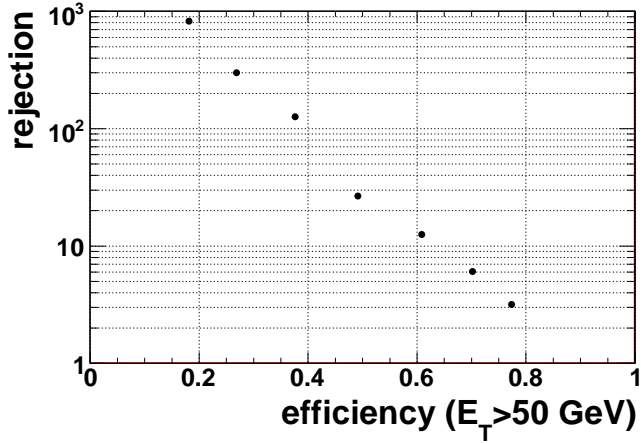}
\caption{Fake jet rejection vs jet reconstruction efficiency. Each point 
represents a different cut in $\sigma_{\Sigma j_T}$ from rejecting 
$\sigma_{\Sigma j_T} >$ -1 (lower right) to $\sigma_{\Sigma j_T} >$ -4 (upper 
left).}
\label{fig:rejvseff}
\end{figure}

\section{Fake jet rejection}\label{sec:fakejets}

Rejection of fake jets is maximized when exploiting the differences between 
the embedded and fake jets. Figure \ref{fig:jet} shows a general feature of 
fake jets by comparing two jets with the same reconstructed jet $E_T$ 
(= 57 GeV). The jet in the left panel is an embedded PYTHIA jet and the jet in 
the right panel is reconstructed from the hard-cut HIJING events. Even though 
the fragmentation of the PYTHIA jet is not uniform in $\eta$ and $\phi$, a 
clear core is seen. The fake jet from the hard-cut HIJING event is a cone 
which stabilized around three towers at large angle to the reconstructed jet 
axis and no core exists. Such a structure is generally indicative of the jets 
from the hard-cut HIJING events: little to no core and energy at large angles 
from the jet axis.

The variable used to discriminate against fake jets is $\Sigma j_{T}$, which 
is constructed to give larger weight to energies at large angle with respect 
to the jet direction. It is defined as
\begin{equation}\label{eq:sumjtdef}
\Sigma j_{T} = \sum_{\mathrm{cell}\in\mathrm{jet}} E_{T,\mathrm{cell}} 
\sin R_{\mathrm{cell}}
\end{equation}
where $E_{T,\mathrm{cell}}$ is the $E_{T}$ with respect to the beam and 
$R_{\mathrm{cell}}$ is the angle (in $\phi$ and $\eta$) of the cell with 
respect to the jet. It has a trivial jet $E_T$ dependence. The definition can 
be rearranged to yield
\begin{equation}\label{eq:sumjtdefrecoet}
\Sigma j_T = \langle \sin R \rangle E_{T,\mathrm{jet}}
\end{equation}
where $\langle \sin R \rangle = \Sigma_{\mathrm{cell}} 
E_{T,\mathrm{cell}}\sin R_{\mathrm{cell}}/E_{T,\mathrm{jet}}$ is the energy 
weighted average of $\sin R$. Therefore, broad distributions would have a 
larger $\Sigma j_T$ than narrow distributions at the same $E_T$. Figure 
\ref{fig:sumjtetdep} shows the jet $E_T$ dependence of $\Sigma j_T$. It is 
linear for the fake jets measured from hard-cut HIJING events. A rising trend 
is also observed in the embedded jets, but, because jets become more 
collimated with increasing $E_T$, $\langle \sin R \rangle$ decreases with 
increasing $E_T$, and, therefore, the rising trend of $\Sigma j_T$ increases 
more slowly at high $E_T$. This has the effect that the discrimination between 
real and fake jets becomes more effective with increasing jet $E_T$.

Because of the $E_T$ dependence of $\Sigma j_T$, the applied cut is an 
$E_T$-independent quantity: the number of RMS from the fake-jet mean, defined 
as
\begin{equation}\label{eq:sigmasumjt}
\sigma_{\Sigma j_T} = \frac{\Sigma j_{T} - \langle \Sigma j_T \rangle 
\left(E_T \right)}{\sigma\left(E_T\right)}
\end{equation}
where $\langle \Sigma j_T \rangle(E_{T})$ and $\sigma\left(E_T\right)$  are 
the $E_T$-dependent mean and RMS, respectively, fitted to the hard-cut HIJING 
jets from Figure \ref{fig:sumjtetdep}. Figure \ref{fig:sigmasumjt} shows the 
distribution of $\sigma_{\Sigma j_T}$ for two different jet $E_T$ bins for 
embedded and fake jets. By definition the fake jets have a mean of zero and an 
RMS of 1 while the embedded jets have an obviously smaller mean in this 
variable. The rate of fakes compared to reconstructed embedded jets is also 
seen in these plots as well as its strong $E_T$ dependence.

Because of the clear separation between the embedded and fake jets, a cut can 
be applied to reject fake jets at a desired level. However, since the 
distributions overlap, there is a trade-off between fake jet rejection and 
efficiency loss. For example, the cut applied in Figure \ref{fig:spectrum} is 
to reject jets with $\sigma_{\Sigma j_T} > $ -2.5. For both $E_T$ bins shown 
in Figure \ref{fig:sigmasumjt} this cut is not only effective in rejecting 
nearly all fake jets but also results in a sizeable loss of real jets. Figure 
\ref{fig:rejvseff} shows the anti-correlation between the rejection of fake 
jets and the reconstruction efficiency of embedded jets. In the end the 
statistics and the desired purity of the data sample will dictate what cut 
will be applied in real data (see Section~\ref{sec:data}).

\section{Results and Discussion}\label{sec:results}

The variable $\Sigma j_T$ has been shown to be effective at rejecting fake, 
reconstructed cone jets from HIJING events. In this section the effect of 
energy loss on the rejection and evaluation of the fake jet rate with real 
data are discussed.

\begin{figure}
\includegraphics[width=\linewidth]{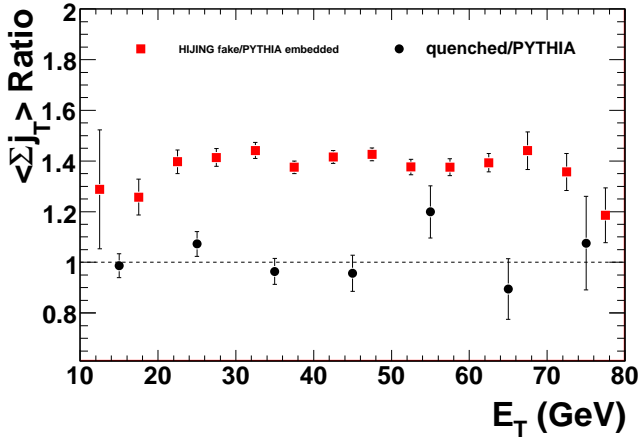}
\caption{The $E_{T}$-dependent ratio of $\langle\Sigma j_{T}\rangle$ in PYQUEN 
jets to embedded jets (black circles) compared to fake HIJING jets to embedded 
jets (red squares) (from Figure \ref{fig:sumjtetdep}). No difference is 
observed between the quenched and unquenched jets and it is much less than the 
difference between the fake and embedded jets.}
\label{fig:pyquenjt}
\end{figure}

\subsection{Effect of energy loss}
In radiative energy loss, the energy profile of a given jet is modified 
jet-by-jet, and on average the energy profile is expected to show 
large modification \cite{Vitev:2008jh}. Because the $\Sigma j_{T}$ is a 
shape cut, it is possible that such a cut could bias results from measuring 
the modification of the energy profile of a jet.

An estimate of the effect can be obtained using Equation 
\ref{eq:sumjtdefrecoet} and assuming a Gaussian distribution of energy within 
a jet. For a Gaussian with an RMS of $\sim$0.2, then $\langle \sin R \rangle$ 
$\sim$ $\langle R \rangle$. At a fixed jet $E_{T}$ the difference between 
embedded and fake $\langle\Sigma j_{T}\rangle$ then is directly related to the 
difference between $\langle R \rangle$. From Figure \ref{fig:sumjtetdep}, at 
50 GeV the fake jet $\langle\Sigma j_{T}\rangle$ is twice that of the embedded 
jet. Therefore, $\langle R \rangle$ of the fake jet is twice that of the 
embedded jet. This is quite a large change for a jet with an RMS of 0.2 to one 
with 0.4. Such a difference between vacuum and medium modified shape may be 
expected for 50 GeV gluon jets \cite{Vitev:2008jh}.

An attempt at studying the energy loss effect on the rejection of 
$\Sigma j_{T}$ has been done by simulating and reconstructing a set of 
PYQUEN \cite{PYQUEN} jets in the ATLAS detector. PYQUEN is a set of routines 
which modifies PYTHIA partons by energy loss and including the radiated gluons 
prior to fragmentation. The events were generated assuming b=0 fm and they 
were not embedded into HIJING. Figure \ref{fig:pyquenjt} shows the ratio of 
$\langle \Sigma j_{T}\rangle$ as a function of $E_T$ ({\it i.e.}~Figure 
\ref{fig:sumjtetdep}) for quenched to unquenched jets (circles) and for 
hard-cut HIJING jets to embedded jets (squares). Little change between the 
quenched and unquenched jets is observed in contrast to the large difference 
between the fake and embedded $\langle\Sigma j_{T}\rangle$.

The jet energy profile in PYQUEN jets are modified, but not to the same extent 
as those predicted in Ref.~\cite{Vitev:2008jh}, leading to much less change in 
$\langle R\rangle$. These results were produced after a simulation through the 
ATLAS calorimeter. It is possible that the segmentation is also coarse enough 
to smear any effect generated by PYQUEN. It should be clear, however, that 
because $\Sigma j_{T}$ is a shape variable, it may show different 
sensitivities to different jet profiles. The final resolution of how sensitive 
must wait for analysis of real data.

\subsection{Rejection in data}\label{sec:data}
Determining the absolute rate of fake jets in real data, regardless of the 
rejection variable, can be done by studying the $\Delta\phi$ distribution 
of the pairs of reconstructed jets. This is a useful variable because 1) the 
position resolution of jets is good, 2) it has a clear signal peak at 
$\Delta\phi = \pi$, and 3) fake jets will be randomly associated with other 
real or fake jets in the events and result in a pedestal. The rejection 
variable distribution should be different for different regions of 
$\Delta\phi$.

\begin{figure}
\includegraphics[width=\linewidth]{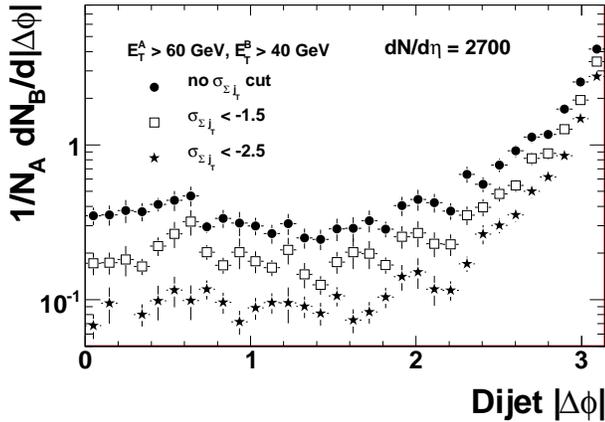}
\caption{$\Delta\phi$ distribution between reconstructed jets with $E_{T}^{A} 
>$ 60 GeV and $E_{T}^{b} >$ 40 GeV for different $\sigma_{\Sigma j_{T}}$ cuts.}
\label{fig:dijetdphi}
\end{figure}

This is the case for $\sigma_{\Sigma j_{T}}$. Figure \ref{fig:dijetdphi} shows 
the distribution of all jets pairs with $E_{T}^{A} >$ 60 GeV correlated with 
with $E_{T}^{B} >$ 40 GeV with different $\sigma_{\Sigma j_{T}}$ cuts applied. 
For the successively hard $\sigma_{\Sigma j_{T}}$ cuts the pedestal is 
suppressed. The stars indicate the cut used in Figure \ref{fig:spectrum} which 
removes a significant fraction of fake jets. With the tightest cut, the signal 
peak at $\Delta\phi = \pi$ becomes slightly narrower. This may indicate that 
the hardest cut removes of the radiated jets in a 3-jet event.

\section{Summary}
\label{sec:summary}
A large fake rate of cone jets from HIJING events is observed, exceeding the 
expected p+p jet rate at low $E_{T}$. They result from stable cones being 
fitted to random and correlated background fluctuations. The variable 
$\Sigma j_{T}$ has been used to identify and reject these fake jets. It is 
shown that this variable is insensitive to the effects of energy loss on jets. 
It is conceivable that neither the extreme fluctuations nor the 
characteristics of fake jets (broader, large angle energy distribution) 
generated in HIJING are observed in the data. In fact, the fake rate will 
strongly depend on the model of the background and, indeed, will require 
measurement in the data itself. Even so, because correlated 
and random fluctuations can produce stable cone jets, it is necessary for any 
jet analysis in heavy ion collisions at RHIC or the LHC to demonstrate a lack 
of contamination of such jets.

\section{Acknowledgements}
\begin{acknowledgement}
This work was partially supported by the Grant Agency of Charles University in
Prague (GA UK 7722/2007).

All results presented use modified versions of ATLAS software and should be 
considered ``ATLAS preliminary''. For completeness, Figures 
\ref{fig:spectrum}-\ref{fig:rejvseff} and \ref{fig:dijetdphi} use ATHENA 
release 12.0.6  while Figure \ref{fig:pyquenjt} uses release 14.2.10.
\end{acknowledgement}


\end{document}